\input harvmac.tex
\vskip 2in
\Title{\vbox{\baselineskip12pt
\hbox to \hsize{\hfill LANDAU-96-TMP-4}
\hbox to \hsize{\hfill}
\hbox to \hsize{\hfill }}}
{\vbox{\centerline{$\kappa$-Symmetry and Bogomol'ny Bound}}}
\centerline{Dimitri Polyakov\footnote{$^\dagger$}
{cftconf@itp.ac.ru,polyakov@pion.rutgers.edu}}
\medskip
\centerline{\it L.D.Landau Institute for Theoretical Physics}
\centerline{\it Kosygina 2, 117940, Moscow V-334, Russia\footnote{$\ddagger$}
{after the 1st of October,1996:IHES,35,Route de Chartres,
91440 Bures-sur-Yvette, France}}
\vskip .5in
\centerline {\bf Abstract}
By applying the twistor-like transform to the GS superstring theory
we fix  the  local
fermionic  gauge symmetry, known as $\kappa$-symmetry.
Fixing the $\kappa$-symetry in the GS superstring theory and 
the $N=1$ worldsheet supersymmetry in the NSR theory,proves                    
the relation between these theories,conjectured earlier;
the GS fermionic variable $\theta$ is shown to be related to the
NSR worldsheet fermionic variable $\psi$ through the twistor
field $\Lambda$.
Upon fixing the $\kappa$-symmetry, we derive the gauge-fixed
expressions for the supercharges (corresponding to spinor
representations of SO(8)) and analyze the gauge-fixed space-time 
superalgebra.The analysis relates the $\kappa$ invariance to
the Bogomol'ny-type condition,with the former appearing as a
 space-time superpartner of the latter.
The BPS saturation appears to be crucial for the non-perturbative
formulation of the $\kappa$-symmetry.
The relation to the RR charges of p-brane solutions of
the $D=10$ supergravity is discussed. 

\Date{August-September 96}
\vfill\eject
\lref\sez{E.Bergshoeff,E.Sezgin,
Nucl.Phys.{\bf B422}(1994)329, hep-th/9312168}
\lref\shenker{D.Friedan,S.Shenker,E.Martinec,Nucl.Phys.{\bf B271}(1986) 93}
\lref\tzeytlin{M.B.Green,J.H.Schwarz,E.Witten,
Superstring Theory,Cambridge University Press (1987)}
\lref\polchino{J.Polchinski,NSF-ITP-95-122,hep-th/9510017}
\lref\sezg{E.Sezgin,{\bf {CTP TAMU-49/95}},hep-th/9512082}
\lref\sezgi{E.Bergshoeff,E.Sezgin,Phys.Lett.{\bf B354}(1995) 256}
\lref\tonin{P.Pasti, M.Tonin, Nucl.Physics {\bf B418}(1994)337}
\lref\green{M.B.Green, Phys.Lett.{\bf B223}(1989)157}
\lref\townsend{P.Townsend,p-Brane Democracy,hep-th/9507048}
\lref\azc{J.A.de Azcarraga, J.P.Gauntlett, J.M.Izquierdo, P.K.Townsend,
Phys.Rev.Lett. {\bf D63} (1989) 2443}
\lref\berk{N.Berkovits,Nucl.Physics B {\bf 379}(1992)96}
\lref\galperin{A.Galperin,E.Sokachev,{\bf BONN-HE-93-05}(1993)}
\lref\berkn{N.Berkovits,Nucl.Phys.{\bf B395}(1993)}
\lref\berkna{N.Berkovits,Nucl.Phys.{\bf B408}(1993)}
\lref\berknat{N.Berkovits,{\bf KCL-TH-94-5}}
\lref\klebanov{S.Gubser,A.Hashimoto,I.R.Klebanov,J.Maldacena
Nucl.Phys.{\bf B472}(1996) 231}
\lref\me{D.Polyakov,Nucl.Phys.{\bf B468}(1996)155}
\lref\myself{D.Polyakov,Nucl.Phys.{\bf B449}(1995)159}
\lref\sezgin{E.Sezgin, Aspects of Kappa Symmetry,hep-th/9310126} 
\lref\volkov{D.P.Sorokin,V.I.Tkach,D.V.Volkov,Mod.Phys.Lett.
{\bf A4}(1989) 901}
\lref\volkovv{I.Bandos,D.Sorokin,D.Volkov,Phys.Lett.{\bf B352}(1995)269}
\lref\volkovvv{I.Bandos,D.Sorokin,M.Tonin,P.Pasti,D.Volkov,
Nucl.Phys.{\bf B446}(1995),79}
\lref\duff{M.J.Duff,R.Khuri,J.X.Lu,String Solitons,Phys.Rept.259 (1995),
213-326}
\lref\kallosh{R.Kallosh,Phys.Lett.{\bf B225}(1989)49}
\lref\witten{E.Witten,Nucl.Phys.{\bf B266}(1986)245}
\centerline{\bf 1.Introduction}
The recent significant achievements in the area of string-string 
dualities  and their both actual and potential 
relevance to the understanding
of superstring theory in general,make the 
study of the dynamics of extended 
objects, such as p-branes ~\refs{\duff}, particularly important.
 The problem of the quantization of p-branes remains unresolved by now,
in particular because of complicated non-linear worldvolume equations of
motion.One problem is that, contrary to the case of string theory,
it is not possible to use the conformal gauge for a worldvolume metric
in case of $p > 1$,which crucially simplifies the worldsheet 
equations of motion in  string theory.Apart from this apparent difficulty,
the known classical formulations of super p-branes are Green-Schwarz ones
rather than NSR.At the same time, even the problem of the covariant
quantization of superstring theory in the GS formulation is still not
solved consistently, in spite of certain important steps in this direction
made in ~\refs{\berkn, \berkna, \berknat, \berk}.
 An important breakthrough in the study of 
extended objects has been made in
 the work by Polchinski, ~\refs{\polchino}
which used the idea of relating the $p>1$ extended
objects to strings with mixed Dirichlet-Neumann boundary conditions
(D-branes).D-branes were shown to break one half of the supersymmetries 
in the theory, i.e. to satisfy the BPS saturation condition.The important
computation made in this paper has shown that these BPS-saturated 
objects carry a complete set of intrinsic Ramond-Ramond electric and
magnetic charges,satisfying the Dirac's quantization condition.
 It should be noted, however, that the D-brane description of
a p-brane dynamics  is the approximate one, true only up to massless
excitations,propagating on the D-brane,which become the collective 
coordinates for the transverse fluctuations. The question
of whether the Dirichlet string - p-brane  correspondence remains true when
the massive modes are taken into account, is unclear and requires the 
knowledge of consistent quantization of p-brane theories.
One particular physical example of  D-brane as a soliton carrying RR 
charges  has been considered in ~\refs{\klebanov}.
\centerline{\bf  2. $\kappa$-symmetry and the twistor-like formalism} 
 Let us now return to the question of the covariant quantization of a GS
superstring, mentioned above.The main obstacle to the quantization is that 
the theory contains a specific constraint relating the 
canonical momenta $P_X$
and $P_\theta$, the canonical conjugates of the target space superpartners
 $X^m$ and $\theta^\mu$ .Because of this constraint, the straightforward
quantization procedure using the Dirac's brackets leads to extremely
cumbersome expressions which cannot be resolved without breaking a general 
covariance. This problem has a very important and subtle connection 
with a specific local fermionic gauge symmetry, known as the 
 $\kappa$ - symmetry ,which  is present in 
the GS formulation of superstring
theory, as well as in super p-brane theories. 
This gauge symmetry, potentially
very important  for the consistent quantization of both GS superstrings
and super p-branes, has a delicate relation to the local worldsheet 
(worldvolume) supersymmetries and plays an important role in insuring the 
supersymmetry in the target space.  
 Namely, in the GS superstring theory it eliminates 8 out of 16 space-time
fermionic degrees of freedom,thus insuring 
that 8 physical bosonic coordinates
 have the equal number of superpartners in the target space.
For a GS superstring with the action ~\refs{\tzeytlin}:
\eqn\grav{\eqalign{S=-{1\over{2\pi}}\int{d^2}\tau{\sqrt{h}}
{h^{ij}}{\Pi_{i}^{m}}{\Pi_{jm}} 
+ {1\over{\pi}}\int{d\tau}(-i{\epsilon^{ij}}
{\partial_i}{X^m}{\theta_\mu}
{\Gamma_m^{\mu\nu}}{\partial_j}{\theta_\nu}),}}
where $\Pi_i^m={\partial_i}{X^m}-i{\theta_\mu}
{\Gamma^m_{\mu\nu}}{\partial_i}{\theta_\nu}$,$m=1,...10$,
$\mu=1,...16$,$i=1,2$;the $\kappa$-transformations
have the form:
\eqn\grav{\eqalign{\delta{\theta^\mu}
=2i{\Gamma^m_{\mu\nu}}\Pi_{im}{\kappa^{i\nu}}\cr
\delta{X^m}=i{\theta^\mu}{\Gamma^m_{\mu\nu}}\delta{\theta^\nu}\cr
\delta({\sqrt{h}}h^{ij})=
-16{\sqrt{h}}{P^{ik}_{-}}{\kappa^{j\mu}}{\partial_k}{\theta_\mu}}} 
with ${P^{ik}_{\pm}}={1\over2}(h^{ik}\pm{{\epsilon^{ik}}\over{{\sqrt{h}}}})$
This may be generalized to a the case of a superstring propagating in a
curved target space, coupled to  SO(32) YM background.
The action then has the form ~\refs{\sez}
\eqn\grav{\eqalign{S=\int{d^2}{\tau}(-{1\over2}{\sqrt{h}}h^{ij}{E^a_i}
{E^a_j}+{1\over2}{\epsilon^{ij}}{\partial_i}{Z^M}{\partial_j}
{Z^N}B_{NM}+\cr+{1\over2}{\psi^I}{\gamma^i}D_i{\psi^J})}}
where $Z^M=(X^m,\theta^\mu)$ denotes the superspace coordinates,
$A=(a,\alpha)$ labels the tangent superspace,
$D_i{\Psi^I}=({\partial_i}{\delta^{IJ}}+{\partial_i}Z^M{A_M^{IJ}}){\Psi^J}$,
the Kalb - Ramond 2-form $B=B_{NM}d{Z^M}{\wedge}d{Z^N}$, $\psi^I$ are
heterotic fermions,$A_M$ is the YM superfield and $E^a_i$ is the pulled back
supervielbein: $E_i^A={\partial_i}Z^M{E^A_M}$.
The $\kappa$-symmetry for this action has the form:
\eqn\grav{\eqalign{\delta{Z^M}=
{\kappa_{i\alpha}}{P^{ij}_{-}}{E^a_j}{\Gamma^{\alpha\beta}_{a}}{E^M_j}\cr
\delta{\psi^I}=-\delta{Z^M}{A_M^{IJ}}{\psi_J}\cr
e_{ir}\delta{e_j^r}={P^m_{+i}}{P^{jn}_{+}}{S_m^\alpha}{\kappa_{n\alpha}}}}
where $S^\alpha_m$ is determined by the $\kappa$-invariance of the action.
It is important that the $\kappa$-invariance condition, along with fixing
$S_m^\alpha$, also imposes the following constraints on the space-time 
supertorsion and YM superfield strengths ~\refs{\sez, \volkovv, \volkovvv}:
\eqn\grav{\eqalign{T^c_{\alpha\beta}=2{\Gamma^c_\alpha\beta}\cr
T_{{\alpha}{\lbrack}bc{\rbrack}}={u^{\beta}_{\lbrack{b}}}
{\Gamma_{c{\rbrack}\beta\alpha}}+{\eta_{bc}}v_{\alpha}\cr
H_{\alpha\beta\gamma}=0, H_{a\alpha\beta}=-2(\Gamma_a)_{\alpha\beta},\cr
H_{ab\alpha}=2(\Gamma_{ab})_{\alpha\beta}v^\beta
+2{u^\beta_{\lbrack{a}}}{\Gamma_{b\rbrack\beta\alpha}}\cr
F_{\alpha\beta}^{IJ}=0,F_{a\alpha}^{IJ}=
(\Gamma_a)_{\alpha\beta}\chi^{\beta{IJ}}\cr
S^\alpha_m=-4e^\alpha_m+2E^a_m(-u^\alpha_a+\Gamma^{\alpha\beta}_a{v_\beta})-
{1\over2}\psi^I{\gamma_i}\psi^J{\chi^\alpha_{IJ}}}}
The generalization to a case of super p-branes is straightforward
(see,for example,~\refs{\sez,\sezgin})
The background constraints required by the $\kappa$-symmetry were shown to
be consistent with those of the pure $N=1$,$D=10$ supergravity,in the absence
of the YM sector, ignoring the $\kappa$-symmetry anomalies ~\refs{\witten}.
 Let us now return to the case of a GS superstring in a $D=10$ flat target 
space and the related problems of quantization.
 As we mentioned above,the $\kappa$-symmetry removes a half of the fermionic
degrees of freedom in the space-time.However, the exact explanation ,showing
how it happens precisely, has not been  given so far.At the same time,
this question is crucial for the gauge fixing in the GS theory.In the
section 4  we will give an answer to that by using the twistor-like
formalism ~\refs{\sez, \volkovv, \volkovvv, \sezgin}.
 The straightforward attempt to fix the gauge by using the standard scheme
does not lead to a complete success because one has to introduce an infinite 
number of ghosts and to deal with regularizing infinite sums;moreover this
approach still leaves a residual gauge symmetry ~\refs{\kallosh}.
These difficulties are related to the fact that there is no conserved 
Noether charge corresponding to the $\kappa$-symmetry and this symmetry is
infinitely reducible,in terms of the BV formalism.
 The alternative approach, known as twistor-like formalism, which we will 
particularly use in this paper,consists of trading the $\kappa$-symmetry
for an extended local worldvolume supersymmetry;the procedure involves
the elevation of a (p+1)-dimensional worldvolume into a superspace with
a necessary number of fermionic components which replace the ones
of the gauge $\kappa$-parameter.The idea is that the resulting action with an 
extended local worldvolume supersymmetry would reproduce the background 
constraints (5) and have the equal number of the gauge parameters.
Such an action may  indeed be constructed and shown to be classically
equivalent to the original one with the $\kappa$-symmetry.This construction
has been performed in details in ~\refs{\sez} for the cases of
 massive superparticles,superstrings and super p-branes.The important element 
of the construction is the twistor-like identity:
\eqn\lowen{{E^{\bar{\alpha}}_{\alpha}}
{\Gamma_{{\bar\alpha}{\bar\beta}}^{\bar{a}}}{E^{\bar\beta}_{\beta}}=
{\Gamma_{\alpha\beta}^a}{E^{\bar{a}}_a}} 
where we have modified our notations: all the indices with the ``bar'' now
refer to the target superspace while those without the bar label the 
superworldvolume.These notations will be used from now on and ,basically,
they follow the reference ~\refs{\sez}.The matrix
 ${E^{\bar{A}}_A}=E^M_A{\partial_M}{Z^{\bar{M}}}{E_{\bar{M}}^{\bar{A}}}$,
where $E^M_A$ and $E^{\bar{M}}_{\bar{A}}$ are supervielbeins in the 
superworldvolume and the target superspace respectively. 
  Particularly, in the superstring case (the case we will be interested in)
the $\kappa$-symmetry contains 8 gauge parameters;therefore, the 
appropriate superstring theory with the $\kappa$-symmetry replaced with
an extended worldsheet supersymmetry,must in fact have the $N=4$ supersymmetry
on the worldsurface (so that the numbers of gauge parameters in both theories
are equal).The appropriate action with the $N=4$ worldsheet supersymmetry is
~\refs{\sez}
\eqn\lowen{S=\int{d^2}\tau{d^8}\theta({P_{\bar{a}}^{\alpha}}+P^{M_1M_2}
({{\tilde{B}}}_{M_1M_2}-{\partial_{M_1}}Q_{M_2}))}
where $P^{\bar{A}}_A$,$P^{M_1M_2}$ and $Q_M$ are Lagrange multiplier
superfields  and the 2-form 
\eqn\grav{\eqalign{{{\tilde{B}}_{M_1M_2}}=
{{(-1)}^{{M_1}(M_2+{\bar{M_2}})}}{\partial_{M_1}}Z^{\bar{M_1}}
{\partial_{M_2}}Z^{{\bar{M_2}}}B_{{\bar{M_1}}{\bar{M_2}}}-\cr
{i\over{32}}{\Gamma^{\alpha\beta}_{c_2}}({E^{c_2}_{M_2}}{E^{c_1}_{M_1}}
H_{\alpha\beta{c_1}}-{E^{c_2}_{M_1}}{E^{c_1}_{M_2}}H_{\alpha\beta{c_1}});\cr
H_{\alpha\beta{c_1}}={E^{\bar{A}}_\alpha}{E^{\bar{B}}_\beta}
{E^{\bar{C}}_{c_1}}H_{\bar{CBA}}}}  
is constructed so that it is closed on the $worldsheet$.
The fermionic superworldsheet index $\alpha$ may be represented as
$\alpha = {\tilde\alpha}r$ where $\tilde\alpha = 1,2$ is the worldsheet
spinor index and $r$ labels the automorphism group of the $N=4,d=2$
supersymmetry;the matrix $\Gamma^a_{\alpha\beta}=
{\gamma^a_{{\tilde\alpha}{\tilde\beta}}}\eta_{rs}$,where 
$\gamma^a$ are 2d gamma-matrices and $\eta$ is
an invariant $4\times4$ tensor of the automorphism group (which is SU(2) in
our case).
Furthermore,the following constraints should be imposed on the worldsurface
 supertorsion:
\eqn\grav{\eqalign{{T^a_{\alpha\beta}}=-2i{(\Gamma^a)_{\alpha\beta}}\cr
T^a_{b\alpha}=0,T^a_{bc}=0,T^\gamma_{\alpha\beta}=0}}
 Let us make the final remark before explaining the idea behind 
the procedure of fixing the $\kappa$-symmetry
gauge.The fact that the $\kappa$-symmetry removes 8 fermionic degrees
of freedom implies that only a half of the total number of supersymmetries
(8 out of 16) actually present in the theory.
This becomes especially clear given the relation between GS and NSR
formulations of superstring theory, discussed in 
~\refs{\myself, \berkn, \berkna, \berknat}
The generator of a space-time supersymmetry in the NSR theory 
~\refs{\shenker}:
$Q_{\bar\mu}=\oint{{dz}\over{2i\pi}}e^{-1/2\phi}\Sigma_{\bar\mu}$,
where $\phi$ is a bosonized superconformal ghost, $\Sigma$ is a spin operator
for matter fields, - has 16 components while the actual number of 
supersymmetries in the space-time is equal to 8.
This leads to the following questions:
1.One may think of the connection between $\kappa$-invariance and BPS
saturation condition,because of the common property of the BPS saturated
extended objects (such as D-branes) and the ones with the $\kappa$-symmetry
 to have a half of the supersymmetries eliminated.
We will show that this relation between the 
$\kappa$-symmetry and the Bogomol'ny bound  does indeed exist,
with the BPS saturation condition being a target-space superpartner of
the $\kappa$-invariance.
 2.The expression for the target space supercharge $Q_{\bar\mu}$ must
be modified upon fixing the $\kappa$-symmetry in order to generate
8 supersymmetries instead of 16. We will obtain the expression
for the gauge-fixed space-time SUSY generator (we will call it the 
``$\kappa$-projected supercharge.
 3.Given the property of the  D-branes (which are BPS saturated 
extended objects)
to carry a set of RR charges,we will discuss the relation of the $\kappa$-
symmetry to the RR charges of the extended objects.
The p-form central charges in the space-time supersymmetry algebra,
related to p-brane solutions of the corresponding low-energy
effective theories,will appear in the O.P.E. between
spin operators for matter fields.
\centerline{\bf 3.Gauge fixing in the NSR superstring theory}
  Before explaining the procedure of fixing the $\kappa$-symmetry gauge
in the GS theory, let us illustrate the method on a much more elementary
example, namely, fixing the gauge symmetry (N=1 worldsheet supersymmetry)
in the NSR theory.In this case fixing the gauge worldsheet supersymmetry
eliminates 2 out of 10 worldsheet fermions,thus leaving the appropriate
number of physical fermionic degrees of freedom on the worldsheet.
The relation implementing the gauge fixing is
\eqn\lowen{S^a=0}
where 
$S^a={1\over2}{\gamma^b}{\gamma^a}{\psi^{\bar{m}}}{\partial_b}X_{\bar{m}}$ is 
the worldsheet supercurrent.
In the components,this condition may be written as
\eqn\lowen{S_{\pm}={\psi^{\bar\mu}_{\pm}}{\partial_\pm}X_{\bar\mu}=0}
Geometrically, this means that ${\psi_{\pm}^{||}}=0$,i.e.
the worldsheet supersymmetry insures that the components of $\psi$
parallel to the worldsurface vanish.Thus the gauge fixing leaves 8 physical
worldsheet fermions out of 10, orthogonal to the worldsheet.
Let us now introduce 10 basic vectors $({\vec{e^a}},{\vec{n^i}})$ in
the space-time,$a=1,2$,$i=1,...,8$ with ${\vec{e^a}}$ tangent and 
$\vec{n^i}$ orthogonal to the worldsurface.We choose:
$({\vec{e^a}},{\vec{e^b}})=\delta_{ab}$,
$({\vec{n^i}},{\vec{n^j}})=\delta_{ij}$,$({\vec{e^a}},{\vec{n^i}})=0$.
The following relations take place:
\eqn\grav{\eqalign{{\psi^{\bar\mu}}=\varphi^a{\vec{e_a}}+
\varphi^i{\vec{n_i}}\cr
\partial_\pm{X^{\bar\mu}}=v^a_\pm{\vec{e^{\bar\mu}_a}}\cr
\partial_\pm{\vec{e^a}}=
{A^{ab}_{\pm}}{\vec{e_b}}+{B^{ai}_\pm}{\vec{n_i}}\cr
\partial_\pm{\vec{n^i}}=C^{ij}_{\pm}{\vec{n_j}}-B^{ai}_{\pm}{\vec{e_a}}}}
Here $v$ is a zweibein,$A,B,C$ are the components of a 10d spin connection
(with $A$ being a 2d spin connection).
Note that, for a superstring propagating in a flat space-time, one may 
choose $B,C=0$.
Fixing the conformal gauge:
$h_{ab}={e^\varphi}\eta_{ab}$ for the worldsheet metric with $\varphi$ being
a conformal factor and $\eta_{ab}$ the 2d Minkowski metric,
one may choose the zweibeins:
$v^1_\pm=\pm{i\over{\sqrt{2}}}e^{{\varphi}\over2}$,
$v^2_\pm={i\over{\sqrt{2}}}e^{{\varphi}\over2}$.
The gauge fixing condition $S^{\pm}=0$ then gives:
\eqn\lowen{{\varphi^a_{\pm}}=\epsilon^{ab}{\varphi_{b\pm}}}
Now, let us substitute this all into the NSR superstring action:
\eqn\lowen{I_{NSR}=\int{d^2}\tau{\eta_{{\bar\mu}{\bar\nu}}}
(\partial{X^{\bar\mu}}{\bar\partial}{X^{\bar\nu}}+
{\psi^{\bar\mu}_{-}}\partial
{\psi^{\bar\nu}_{-}}+{\psi^{\bar\mu}_{+}}
{\bar\partial}{\psi^{\bar\nu}_{+}})}
where $\eta_{{\bar\mu}{\bar\nu}}$ is the Minkowski metric in a 10d
target space with the signature :$(-+,+...+)$.
The second term in the $I_{NSR}$ becomes:
\eqn\grav{\eqalign{{\psi^{\bar\mu}_{-}}\partial{\psi^{\bar\nu}_{-}}
{\eta_{{\bar\mu}{\bar\nu}}}=
{\varphi^1_{-}}{\varphi^2_{-}}({A^{12}_{-}}+{A^{21}_{-}})+\cr
{\varphi^i_{-}}{\varphi^a_{-}}({B^{ai}_{-}}-{B^{ai}_{-}})+{\varphi^i_{-}}
{C_{ii}^{-}}{\varphi^i_{-}}+\cr
({\varphi^1_{-}}{\partial}{\varphi^1_{-}}-{\varphi^2_{-}}
\partial{\varphi^2_{-}})+{\varphi^i_{-}}\partial{\varphi^i_{-}}}}
The first three terms in this expression vanish automatically the 4th and
5th cancel due to (18) and only the last term remains.
Analogously,the constraint $S^{+}=0$ leads to
\eqn\lowen{{\psi^{\bar\mu}_{+}}{\bar\partial}{\psi_{{\bar\mu}+}}=
{\varphi^i_{+}}{\bar\partial}{\varphi^i_{+}}.}
Apart from that, the reparametrizational invariance condition meaning
$T=0$, $T$ is the stress-energy tensor, removes 2 out of 10 bosons from
the theory, leaving only the transverse ones.Therefore, the gauge-fixed NSR 
action :
\eqn\lowen{{I_{NSR-g.f.}}=\int{d^2}\tau(\partial{X^i}{\bar\partial}{X^i}+
{\varphi^i_{+}}{\bar\partial}{\varphi^i_{+}}+
{\varphi^i_{-}}\partial{\varphi^i_{-}})}
consists of the ``transverse'' worldsheet bosons and fermions only.
Next, we are going to generalize the method developed in this section
 to fix the gauge in the NSR theory, to the case of the $\kappa$-symmetry,
in order to fix the gauge in the GS action.The important step will be to
use the twistor-like formalism  replacing the GS action with the
classically equivalent $N=4$ worldsheet supersymmetric action (6)(?) and
to find the corresponding 4 worldsheet supercharges $Q^r_{\pm}$,$r=1,...,4$.
Then,the gauge fixing will be determined by the constraints
\eqn\lowen{Q^r_{\pm}=0,}
classically equivalent to the $\kappa$-invariance condition.
These constraints will eliminate 8 out of 16 
components of $\theta^{\bar\mu}$,
leaving the ``gauge-fixed'' GS fermionic variable $\theta^i,i=1,...8$.
 We will see the remarkable relation of this variable to the orthogonal
worldsheet fermions $\varphi^i$ in the NSR formalism, thus coming to the 
conclusion that the relation between NSR and GS formulations of superstring
theory is insured by the $\kappa$-symmetry.
\centerline{\bf 4. Gauging away the $\kappa$-symmetry}
 Let us consider the action (7) and the associated supertorsion
 constraints (9) on the worldsheet.The local $N=4$ supersymmetry
transformations, the part of
 the worldsheet superdiffeomorphisms, preserving the constraints (9) and
leaving the action (7) invariant are given by:
\eqn\grav{\eqalign{\delta{\tau^a}=-{i\over2}{\gamma^a_{{\tilde\alpha}
{\tilde\beta}}}{\eta^{rs}}{\theta_{{\tilde\alpha}r}}
{D^{\tilde\beta}_s}\lambda \cr
\delta{\theta^r_{\tilde\alpha}}=-{i\over2}{D^r_{\tilde\alpha}}\lambda}}
Here $\tau^a$ are  worldsheet coordinates,$\gamma^a$ are 2d gamma-matrices,
$\eta^{rs}$ is a $4\times{4}$ SU(2) invariant tensor,$\lambda$ is an
arbitrary infinitezimal superfield.
This is the 2d generalization of the worldline superdiffeomorphisms
of ~\refs{\sez}:
The equations of motion  corresponding to the action (7) are given by:
\eqn\grav{\eqalign{{E^a_{{\tilde\alpha}r}}=0\cr
{\partial_{M_1}}{{\tilde{B}}_{{M_2}{M_3}}}+cyclic{\lbrack}{M_1M_2M_3}
{\rbrack}={{\tilde{H}}_{M_1M_2M_3M_4}}=0\cr
{\partial_{M_1}}P^{M_1M_2}=0}}
The straightforward computation,using (7),(19),taking into account
(20) and transformation properties of supertensors gives the following
expression for the supercharges:
\eqn\lowen{Q^r_{\tilde\alpha}=\int{d^2}\tau{d^8}\theta
({\gamma^a_{{\tilde\alpha}
{\tilde\beta}}}{\eta^{rs}}{P^{\tilde\beta}_{{\bar{a}}s}}{E_a^{\bar{a}}})}
Analogously to (10), the $\kappa$-invariance condition, written in terms
of the $N=4$ worldsheet supersymmetry,is given by
\eqn\lowen{{S^r_{\tilde\alpha}}=\int{d^8}\theta
{\gamma^a_{{\tilde\alpha}{\tilde\beta}}}\eta^{rs}
{P^{\tilde\beta}_{{\bar{a}}s}}{E^{\bar{a}}_a}=0}
Similarly to the NSR case, this becomes the gauge-fixing condition
for the GS superstring theory, which gauges away the $\kappa$-symmetry.
Here (22) is the system of 8 equations with 16 variables,i.e.
the one that gauges away  8 space-time fermionic components.
We will now concentrate on analyzing (22) for a flat target superspace,
in order to find the gauge-fixed expression for the $\theta^{\bar\mu}$.
The space-time supervielbeins may be chosen as follows:
\eqn\grav{\eqalign{{E^{\bar{a}}_{\bar{m}}}={\delta^{\bar{a}}_{\bar{m}}},
{E^{\bar\alpha}_{\bar{m}}}=0,{E^{\bar\alpha}_{\bar\mu}}=
{\delta^{\bar\alpha}_{\bar\mu}}\cr
{E^{\bar{a}}_{\bar\mu}}={\Gamma^{\bar{a}}_{{\bar\mu}{\bar\nu}}}
{\theta^{\bar\nu}}}}
and the expansions of the space-time supercoordinates
$Z^{\bar{M}}=(X^{\bar{m}},{\theta^{\bar\mu}})$ in 
$\theta_{\tilde\alpha}^r$ is
given by:
\eqn\grav{\eqalign{X^{\bar{m}}(\tau^a,{\theta_{\tilde\alpha}^r})=
X^{\bar{m}}(\tau^a)+{\theta^r_{\tilde\alpha}}
{\psi^{{\bar{m}}{\tilde\alpha}}_r}
+...\cr
\theta^{\bar\mu}(\tau^a,{\theta^r_{\tilde\alpha}})=
{\theta^{\bar\mu}}(\tau^a)+
{\theta^r_{\tilde\alpha}}{\Lambda^{{\bar\mu}{\tilde\alpha}}_r}+...}}
where $\Lambda$ is a twistor-like variable,a commuting worldsheet
and space-time spinor.
Using (23),(24), the expression for the matrix $E^{\bar{A}}_A$ and 
performing the integration over $\theta^r_{\tilde\alpha}$
we obtain after the computation:
\eqn\lowen{{\gamma^a_{{\tilde\alpha}{\tilde\beta}}}{\eta^{rs}}{p_{\bar{a}}}
{\lbrack}{\partial_a}{\psi^{\bar{a}}_{{\tilde\beta}s}}+{\partial_a}
{\theta_{\bar\mu}}{\Gamma^{\bar{a}}_{{\bar\mu}{\bar\nu}}}
{\Lambda_{\bar\nu}^{{\tilde\beta}s}}+{\partial_a}
{\Lambda^{{\tilde\beta}s}_{\bar\mu}}{\Gamma^{\bar{a}}_{{\bar\mu}{\bar\nu}}}
{\theta_{\bar\nu}}{\rbrack}=0}
where
\eqn\lowen{p_{\bar{a}}={{(D_{{\tilde\alpha}s})}^7}
{P_{\bar{a}}^{{\tilde\alpha}s}}{|_{\theta=0}}}
Several conclusions may now be drawn from the equation (25),
which plays the central role in fixing the gauge.
 Remarkably, it relates the NSR-like variable $\psi$ to 
the GS-like variable $\theta$,through the twistor-like field $\Lambda$.
As we will see,the worldsheet fermions $\psi$'s in (25) are
related to the ``transverse'' fermions $\varphi^i$,
introduced earlier in the NSR formalism.
 The relation between the NSR and GS formulations of superstring theory,
which has been discussed earlier in a number of papers, including
~\refs{\berknat, \berkna, \berkn, \myself},
 arises as a result of fixing the $\kappa$ -symmetry
gauge in the GS theory and the $N=1$ worldsheet supersymmetry in the 
NSR formulation.Furthermore, as we will show, the equation (25)
leads, at the same time, to some additional constraints, related
to Bogomol'ny-type conditions.
Consider now the first term in (25),related to the
``worldsheet part'' of the supercurrent ${S_{\tilde\alpha}^r}$.
Let us denote
\eqn\lowen{{\partial_a}{\gamma^a_{{\tilde\alpha}{\tilde\beta}}}
\psi^{{\bar{a}}a}_{{\tilde\beta}r}
={\partial_a}{{\tilde\psi}^{{\bar{a}}a}_{{\tilde\alpha}r}}.}
Note that the Lagrange multiplier field $p_{\bar{a}}$ does not in
fact describe any new degree of freedom, as was pointed out in 
~\refs{\galperin, \sez}.Moreover,from  the O.P.E arguments:
$S(z)S(w)\sim{{const}\over{{(z-w)}^3}}+...$ 
and $\partial\psi(z)\partial\psi(w)\sim{{const}\over{{(z-w)}^3}}$
we deduce $p(z)p(w)\sim{const}+...$ or, in other words,
the Lagrange multiplier field $p_{\bar{a}}$ is of conformal dimension 0.
Without a loss of generality, we may therefore choose $p_{\bar{a}}$
to be a constant (on the worldsheet) 10-vector.Then,
let us further introduce a new fermionic field $\tilde\varphi$:
\eqn\lowen{p_{\bar{a}}{\partial_a}
{{\tilde\psi}^{{\bar{a}}a}_{{\tilde\alpha}r}}
{{=}^{def}}{\partial_a}{{\tilde\varphi^a}_{{\tilde\alpha}r}}}
The indices $\tilde\alpha=1,2$ and $r=1,...,4$ may be unified into one
8-dimensional index $i=1,...,8$;then the first term of (25) may be written as
\eqn\lowen{{\gamma^a_{{\tilde\alpha}{\tilde\beta}}}{\eta_{rs}}p_{\bar{a}}
{\psi^{\bar{a}}_{{\tilde\beta}s}}={\partial_+}{{\tilde\varphi}_i^{-}}+
{\partial_-}{{\tilde\varphi}_i^{+}}.}
We will show now that the index $i$ corresponds to 8-dimensional 
(namely, the vector)
representation of SO(8), i.e. it is SO(8) covariant.  
Indeed, the  SU(2) invariant metric ${\eta_{rs}}$ may be chosen as
\eqn\lowen{\eta={I_{2\times{2}}}\otimes{\left[\matrix{0&1\cr{-1}&0}\right]}}  
where $I_{2\times{2}}$ is the $2\times{2}$ unit matrix
and the metric spinor alternating the index $\tilde\alpha$ is given by
${\left[\matrix{0&1\cr{-1}&0}\right]}$.
Therefore the metric tensor alternating the index $i$ may be chosen
 as 
\eqn\lowen{{\eta_{8\times{8}}}={I_{2\times{2}}}\otimes
{\left[\matrix{0&1\cr{-1}&0}\right]}\otimes
{\left[\matrix{0&1\cr{-1}&0}\right]}}
which, after the appropriate 2d rotation, becomes
\eqn\lowen{{I_{2\times{2}}}\otimes{\left[\matrix{i&0\cr{0}&{-i}}\right]}
\otimes{\left[\matrix{i&0\cr{0}&{-i}}\right]}=-{I_{2\times{2}}}\otimes
{\left[\matrix{1&0\cr{0}&{-1}}\right]}\otimes
{\left[\matrix{1&0\cr{0}&{-1}}\right]}}
We see that the index $i$ is SO(4,4) covariant or, after the appropriate 
Wick's rotations, SO(8) covariant.
Finally, let us redefine:
\eqn\grav{\eqalign{{{\tilde\varphi}_i^{\pm}}={\varphi_i^{\pm}},i=1,...,4\cr
{{\tilde\varphi}_{i}^{\pm}}=i{\varphi_i^{\pm}},i=5,...,8}} 
Note that the worldsheet fermions $\varphi^i$ have the same worldsheet
properties (conformal dimension ${1\over2}$) and transform according
to the same 8-dimensional representation of SO(8) as the ``transverse''
NSR fermions $\varphi^i$,introduced in the previous section.
One should therefore identify them and indeed,
 as we will see shortly, 
such an identification insures that the l.h.s. of (25) vanishes,
as required by the $\kappa$-symmetry.
Let us consider next the second and the third terms of (25).
Using the same arguments as above,we redefine:
\eqn\grav{\eqalign{\eta_{rs}{\gamma^a_{{\tilde\alpha}{\tilde\beta}}}
{\Lambda^{{\tilde\beta}s}_{\bar\nu}}{{=}^{def}}
{{\tilde\Lambda}^{{\tilde\alpha}ra}_{\bar{\nu}}}=
{{\tilde\Lambda}^{ia}_{\bar{\nu}}};\cr
{\tilde\Lambda}^{i}=\Lambda^i,i=1,...,4\cr
{\tilde\Lambda}^{i}=i\Lambda^i,i=5,...,8}}
and the equation (25) becomes:
\eqn\grav{\eqalign{{\partial_+}{\varphi^i_{-}}+{\partial_-}{\varphi^i_{+}}
+{\partial_-}\theta_{\bar{\mu}}(\Gamma^{{\bar\mu}{\bar\nu}})
{\Lambda_{\bar\nu}^{i+}}+{\partial_+}{\theta_{\bar\mu}}
(\Gamma^{{\bar\mu}{\bar\nu}}){\Lambda_{\bar\nu}}+\cr
+{\partial_+}{\Lambda_{\bar\mu}^{i-}}({\Gamma}^{{\bar\mu}{\bar\nu}})
\theta_{\bar{\nu}}+{\partial_-}{\Lambda^{i+}_{\bar\mu}}=0}}
where
$\Gamma_{{\bar\mu}{\bar\nu}}=
p_{\bar{a}}\Gamma^{\bar{a}}_{{\bar\mu}{\bar\nu}}$.
In order for the l.h.s. of (25) to vanish we must then have
\eqn\lowen{{\partial_{\pm}}{\theta_{\bar\mu}}({\Gamma_{{\bar\mu}{\bar\nu}}})
{\Lambda_{\bar\nu}^{i\mp}}+{\partial_\pm}{\Lambda_{\bar\mu}^{i\mp}}
\theta_{\bar\nu}=const{\partial_\pm}{\varphi^i_\mp}}
Given that $\varphi^i$'s of this equation are identified with the
``transverse'' worldsheet fermions of (17),the $\kappa$-symmetry
condition (25) will then be satisfied due to the equations of motion for
the $\varphi^i$'s:$\partial_{\pm}{\varphi^i_{\mp}}=0$. 

Thus, we have found that fixing the local fermionic gauge $\kappa$-symmetry
in the GS formulation of superstring theory yields the action (17),
along with the constraint (36) relating the GS and NSR variables.
The fulfillment of (36) leads to some extra conditions which we will
analyze in the next section.We will see their relation to the
BPS saturation criteria.
\centerline{\bf 5.NSR-GS relation and p-form charges}
In the paper ~\refs{\myself}
we  analyzed the relation between 
the NSR and GS formulations of superstring theory in a flat 10d target
space.
Our conclusion,based on the analysis of the worldsheet and target space 
properties of the GS space-time fermionic variable $\theta^{\bar\mu}$
was that it has to be related to the NSR theory through the relation:
\eqn\lowen{\theta^{\bar\mu}=e^{\phi\over2}\Sigma^{\bar\mu}}
where $\Sigma^{\bar\mu}$ is the 10d spin operator for matter fields.
We have shown that, up to a picture-changing,the relation (37) transforms
the GS stress-energy tensor into the stress-energy tensor of the NSR
formulation; also we have shown (37) to be consistent with the GS
equations of motion, in terms of the O.P.E. in the NSR formalism;
and to reproduce the space-time supersymmetry transformations
of the GS theory, up to the operation of picture-changing.
The important expressions we used to prove these relations
were the one for the conjugate momentum :
$P_\theta^{\bar\mu}={i\over{\pi}}e^{\phi\over2}\Sigma^{\bar\mu}$;
and the one for the space-time supersymmetry generator 
~\refs{\shenker} in the 
$+{1\over2}$-picture: 
\eqn\lowen{:\Gamma_1\oint{{dz}\over{2i\pi}}e^{-{\phi\over2}}
\Sigma_{\bar\mu}:=
{-{1\over2}}\oint{{dz}\over{2i\pi}}{\theta_{\bar\nu}}
{\Gamma^{\bar{m}}_{{\bar\mu}{\bar\nu}}}{\partial}X_{\bar{m}}}
where $\Gamma_1=:e^{\phi}(S_{matter}+S_{ghost}):$ is the picture-changing 
operator:
\eqn\lowen{:\Gamma_1:=-{1\over{2}}e^{\phi}{\psi^{\bar\mu}}{\partial}
{X_{\bar\mu}}-{1\over2}e^{2\phi-\chi}{\partial\phi}b+e^{\chi}c}
Analyzing the $\kappa$-symmetry in the GS superstring theory and
exploring the NSR - GS correspondence,
we deduced the expression for the local
NSR operator generating the transformations which
, given the NSR-GS
connection,correspond to the $\kappa$-transformations in the GS formulation:
\eqn\lowen{G^{\bar\mu}_{\kappa}=e^{-{3\over2}\phi}{\Sigma^{\bar\mu}}}
Note that both the well-known expression for the space-time supercharge 
and for the $\kappa$-symmetry generator in the NSR picture,correspond
to the situation when the gauge freedom 
corresponding to the $\kappa$-symmetry
is not yet fixed: they both have 16 components while the gauge-fixed
 space-time supersymmetry (as well as the $kappa$-symmetry) should involve
8 degrees of freedom only.In order to obtain the proper
expressions for these generators in the gauge-fixed theory
one should consider the $\kappa$-projections,
similar to the one performed with $\theta^{\bar\mu}$ in (25).
To explore this question in details,
let us first consider the NSR formulation of the twistor field 
$\Lambda^{i\pm}_{\bar\mu}$ of (35),(36), analogous to the NSR
formulation of the space-time GS fermion $\theta_{\bar\mu}$,
which is the worldsheet superpartner of $\Lambda$.
Before doing it, it would be instructive to find the 
worldsheet superpartner of the field $e^{\phi\over2}$ in the
regular NSR theory with the $N=1$ worldsheet supersymmetry.
 The found expression may then be easily generalized to a case 
of the $N=4$ worldsheet supersymmetry relevant to our approach.
 The expression to be found is relevant to the NSR formulation of
the doubly supersymmetric
approach of ~\refs{\volkovv, \volkovvv}
(the case of N=1 SUSY both on the worldsheet and
in the target space)
  To procede with our computations,we will choose to use the
picture-changed version of the $N=1$ worldsheet  supercharge:
\eqn\lowen{:{\Gamma_1}S^{\pm}:=:{\Gamma_1}\oint{{dz}\over{2i\pi}}
(-{1\over2}{\psi^{\bar\mu}}{\partial}{X_{\bar\mu}}-{1\over2}b\gamma+
c\partial\beta+{3\over2}\beta\partial{c}).}
Applying this picture-changed worldsheet supercharge to the GS
fermionic field $\theta^{\bar\mu}={e^{\phi\over2}}\Sigma^{\bar\mu}$,
we obtain after lengthy computations:
\eqn\grav{\eqalign{\Lambda^{\pm}_{\bar\mu}=:\oint{{dz}\over{2i\pi}}
:{\Gamma_1}S^{\pm}:(w):\theta_{\bar\mu}:(z)=\oint{{dz}\over{2i\pi}}
{1\over{z-w}}({1\over4}e^{{5\over2}\phi-\chi}b\Sigma_{\bar\mu}
S^{\pm}_{matter})+\cr+{\lbrace}Q_{BRST},...{\rbrace}=
-{1\over4}e^{{5\over2}\phi-\chi}b\Sigma_{\bar\mu}S_{matter}^{\pm}}}
 The generalization of this expression to our $N=4$ case is quite 
straightforward, given that the necessary condition (36) for the
$\kappa$-symmetry is fulfilled.We find that, up to a constant, 
 related to the constant in (36),which will be identified later,
 the NSR formulation of
the twistor field $\Lambda^{i}_{\bar\mu}$ is given by:
\eqn\lowen{\Lambda^{i\pm}=
(const) e^{{{5\phi}\over2}-\chi}b{\Sigma_{\bar\mu}}
{\partial_\mp}\varphi^i_{\pm}}
In order to show that this NSR formulation 
of the twistor field is consistent
with the $\kappa$-invariance , we have to show that it satisfies (36)
(up to picture-changing) in terms of the O.P.E. of the NSR formalism.
Thus, we have: 
\eqn\grav{\eqalign{\partial{\theta_{\bar\mu}}\Gamma^{{\bar\mu}{\bar\nu}}
{\Lambda^{i-}_{\bar\nu}}+{\partial}{\Lambda^{i-}_{\bar\mu}}
{\Gamma^{{\bar\mu}{\bar\nu}}}{\theta_{\bar\nu}}=\cr=(const)
e^{3\phi-\chi}{\lbrace}b\partial{\varphi^{i-}}{\Gamma^{{{\bar{m}}_1}...
{{\bar{m}}_7}}}{\psi_{{\bar{m}_1}}}...{\psi_{{\bar{m}}_7}}+\cr+
({\partial}b\partial{\varphi^{i-}}-{1\over2}{\partial^2}{\varphi^{i-}}+
{1\over2}\partial\phi\partial{\varphi^{i-}}){\Gamma^{{{\bar{m}}_1}...
{{\bar{m}}_5}}}{\psi_{{\bar{m}}_1}}...{\psi_{{\bar{m}}_5}}+\cr+
P^{(3)}(\partial{b},\partial\phi,\partial{\varphi^{i-}})
{\Gamma^{{{\bar{m}}_1}...{{\bar{m}}_3}}}{\psi_{{\bar{m}}_1}}...
{\psi_{{\bar{m}}_3}}+{P^{(4)}}(\partial{b},\partial\phi
,\partial{\varphi^{i-}}){\Gamma^{\bar{m}}}{\psi_{\bar{m}}}{\rbrace}+\cr
e^{3\phi-\chi}{\Gamma^i}{\lbrace}b{\Gamma^{{{\bar{m}}_1}...{{\bar{m}}_{10}}}}
{\psi_{{\bar{m}}_1}}...{\psi_{{\bar{m}}_{10}}}+
(\partial{b}-{1\over2}\partial\phi)
{\Gamma^{{{\bar{m}}_1}...{{\bar{m}}_8}}}{\psi_{{\bar{m}}_1}}...
{\psi_{{\bar{m}}_8}}+\cr+
({1\over2}\partial^2{b}-{1\over4}\partial^2{\phi}+{1\over2}\partial\phi
\partial{b}){\Gamma^{{{\bar{m}}_1}...{{\bar{m}}_6}}}
{\psi_{{\bar{m}}_1}}...{\psi_{{\bar{m}}_6}}+\cr+
{G^{(3)}}(\partial{b},\partial\phi){\Gamma^{{{\bar{m}}_1}...{{\bar{m}}_4}}}
+{G^{(4)}}(\partial{b},\partial\phi){\Gamma^{{{\bar{m}}_1}{{\bar{m}}_2}}}
{\rbrace}+e^{3\phi-\chi}{\varphi^{i-}}b
{\Gamma^{{{\bar{m}}_1}...{{\bar{m}}_9}}}{\psi_{{\bar{m}_1}...{\bar{m}}_9}}}}
Here $\Gamma^i$ stands for $\Gamma^{\bar{m}}{\vec{n^i_{\bar{m}}}}$,
$\Gamma^{{{\bar{m}}_1}...{{\bar{m}}_k}}$ denotes the antisymmetrized
product of $k$  10d gamma-matrices contracted with 
$\Gamma_{{\bar\mu}{\bar\nu}}$; and
$P^{(j)}(\partial{b},\partial\phi,\partial{\varphi^{i-}})$ with
$G^{(j)}(\partial{b},\partial\phi)$ are the certain polynomials in
the derivatives of the fields $b$,$\phi$ and $\varphi$,
such that all the terms these polynomials consist of,
have the same conformal dimension $j$ (the example of such a term is
$\partial^2{b}\partial^{j-2}{\phi}$).These polynomials appear in
the process of computing the O.P.E. between $\theta$ and $\Lambda$
 $j$ (see also ~\refs{\myself}).The
precise expressions for them are not given here because
of their length;however, one may show that the lengthy expression
(44) for the operator product may be rewritten in the 
much more convenient form (from now on we will suppress
the signs $+,-$ in the indices for the sake of brevity):
\eqn\grav{\eqalign{\partial{\theta_{\bar\mu}}{\Gamma^{{\bar\mu}{\bar\nu}}}
{\Lambda^{i}_{\bar\nu}}+{\partial}{\Lambda^{i}_{\bar\mu}}
{\Gamma^{{\bar\mu}{\bar\nu}}}{\theta_{\bar\nu}}=\cr
=const{\lbrace}{\Gamma_2}(\partial{\varphi^{i}}+({\Gamma^{{{\bar{m}}_1}
{{\bar{m}}_2}}}{\psi_{{\bar{m}}_1}}{\psi_{{\bar{m}}_2}}+
{\Gamma^{\bar\mu}}\partial{X_{\bar\mu}}){\varphi^{i}})+\cr+
{\Gamma_1}e^{\phi}({\Gamma^{{{\bar{m}}_1}...{{\bar{m}}_5}}}
{\psi_{{\bar{m}}_1}}...{\psi_{{\bar{m}}_5}}{\varphi^{i}}+
{\Gamma^{{{\bar{m}}_1}...{{\bar{m}}_4}}}{\psi_{{\bar{m}}_1}}...
{\psi_{{\bar{m}}_4}}+{\Gamma^{{{\bar{m}_1}}...{{\bar{m}}_3}}}
{\psi_{{\bar{m}}_1}}...{\psi_{{\bar{m}}_3}}\partial{\varphi^i})\cr
+e^{2\phi}({\Gamma^{{{\bar{m}}_1}...{{\bar{m}}_9}}}
{\psi_{{\bar{m}}_1}}...{\psi_{{\bar{m}}_9}}\partial{X^i}+
{\Gamma^{{{\bar{m}}_1}...{{\bar{m}}_8}}}{\psi_{{\bar{m}}_1}}...
{\psi_{{\bar{m}}_8}}\partial{\varphi^i}+\cr+
{\Gamma^{{{\bar{m}}_1}...{{\bar{m}}_7}}}{\psi_{{\bar{m}}_1}}...
{\psi_{{\bar{m}}_7}}({\varphi^j}\partial{X_j}){\varphi^i}+
{\Gamma^{{{\bar{m}}_1}...{{\bar{m}}_6}}}{\psi_{{\bar{m}}_1}}...
{\psi_{{\bar{m}}_6}}({\varphi^j}\partial{X_j})\partial{X^i}){\rbrace}}}
Here $\Gamma_1$ and $\Gamma_2 =:\Gamma_1\Gamma_1:$ are picture-changing
operators. 
In order to check that this expression is indeed consistent with the
$\kappa$-invariance condition (25) we have to show that the entire
expression  (45) is proportional (up to picture-changing) 
to $\partial\varphi^i$ ; or,in other words,that the p-form terms 
(with $p=1,...9$) that
occured in (45) do vanish somehow.
 In the next section we will see, from the analysis of the 
extended space-time 
superalgebras, that the conditions, necessary for the vanishing
of these terms,
are closely related to the Bogomol'ny-type criteria and the BPS
saturation of the corresponding p-brane solutions to the low-energy
effective theory.Furthermore, as we will show, 
the Bogomol'ny condition itself 
will appear to be the target space superpartner of the
$\kappa$-invariance condition (25).
\centerline{\bf 6.New space-time superalgebras and the BPS condition}
In the paper ~\refs{\green} the following extended
space-time superalgebra has been proposed for the GS superstring
theory:
\eqn\grav{\eqalign{{\lbrace}{Q_{\bar\mu}},Q_{\bar\nu}{\rbrace}=
{\Gamma^{\bar{m}}_{{\bar\mu}{\bar\nu}}}P_{\bar{m}}\cr
{\lbrack}P_{\bar{m}},Q_{\bar\mu}{\rbrack}=
-{{(\Gamma_{\bar{m}})}_{{\bar\mu}{\bar\nu}}}T^{\bar\nu}}}
where the new fermionic generator $T^{\bar\nu}$ has been introduced.
 In the references ~\refs{\sezgi, \sezg}
  the generalizations of this 
superalgebra have been found for supermembranes and for
super p-brane theories.
It is not difficult to show now that 
the generator $T^{\bar\nu}$ is related in fact to the 
NSR formulation of the $\kappa$-symmetry generator.
Using the NSR expressions for the space-time supercharge:
$Q_{\bar\mu}=\oint{{dz}\over{2i\pi}}e^{-\phi\over2}\Sigma_{\bar\mu}$
and for the momentum in the $-1$-picture:
$P_{\bar{m}}=e^{-\phi}{\psi_{\bar{m}}}$ we compute the commutator
\eqn\lowen{{\lbrack}P_{\bar{m}},Q_{\bar\mu}{\rbrack}=
-\oint{{dz}\over{2i\pi}}(:e^{-\phi\over2}\Sigma_{\bar\mu}:(z)
:e^{-\phi}\psi^{\bar{m}}:(w))={\Gamma^{\bar{m}}_{{\bar\mu}{\bar\nu}}}
e^{-{3\over2}\phi}\Sigma_{\bar\nu}(w)}
We see that the result, $T^{\bar\nu}$ coincides with the expression
for the $\kappa$-generator in the NSR formalism, introduced earlier.
Though we would like to stress once again
that  the NSR expressions we used in (47) refer to the situation
$before$ fixing the $\kappa$-symmetry gauge, 
we can already make some important conjectures about the relation between
the $\kappa$-invariance and the BPS condition.
Thus,it has been shown in ~\refs{\azc, \townsend} that the 
expression for the anticommutator of two supercharges in (46)
may also be modified to include p-form charges, which have 
been shown to correspond to p-brane solutions of the low-energy
effective theory. 
 In particular, in ~\refs{\townsend}
 the following extension for the superalgebra has been considered:
\eqn\lowen{{\lbrace}Q_{\bar\mu},Q_{\bar\nu}{\rbrace}=
{\Gamma^{\bar{m}}_{{\bar\mu}{\bar\nu}}}(P_{\bar{m}}+T_{\bar{m}})+
{\Gamma^{{{\bar{m}}_1}...{{\bar{m}}_5}}}Z_{{{\bar{m}}_1}...{{\bar{m}}_5}}}
The self-dual 5-form charge $Z$ has appeared to be the central charge 
in the $N=1$,$D=10$ supertranslation algebra; one could then relate
the appropriate magnetic charge ( corresponding to the magnetic-type
non-singular 5-brane solution) to a contraction of a 5-form $Z$
with the outer product of 5 translation Killing vectors of the
fivebrane solution.The 1-form T corresponded to the 1-form electric charge 
carried by the singular elementary string solution.
Generalizations of this superalgebra corresponding to other
p-brane solutions of the $N=1$ $D=10$ supergravity have also been 
considered in these papers.
Returning to the space-time superalgebra (46) we see that
 $T^{\bar\nu}$, identified as the NSR-formulated generator of the 
$\kappa$-transformations may be viewed
as the superpartner of the momentum generator in (46),
supplemented with $p$-form charges;while
 the $\kappa$-invariance condition (which may be expressed naively as
$T^{\bar\nu}=0$),would then correspond to  the vanishing of this
anticommutator,i.e. to the Bogomol'ny-type condition. 
 Nevertheless, in order to address the problem more accurately,
and to develop the above intuitive arguments, one has to
fix the $\kappa$-symmetry gauge  first.Upon fixing the gauge,
the space-time supersymmetry generator $Q_{\bar\mu}$ with 16 components
will have to be replaced with certain gauge-fixed 8-component supersymmetry
 while the generator of the $\kappa$-transformations
will enter the gauge-fixed superalgebra as in (25).
We will argue that the p-form terms that occured in the NSR
expressions (44),(45) for the $\kappa$-symmetry generator, rewritten
in the twistor-like formalism, are related to the p-form charges
in the anticommutator of the extended space-time superalgebra (48)(?).
The Bogomol'ny-type constraint will then insure that the p-form terms
of (45) will cancel,i.e. the $\kappa$-invariance condition will be
fulfilled. 
To show that, we have to deduce first the expression for the supercharge
$Q^i,i=1,...8$ since the 16-component NSR formulation $Q_{\bar\mu}$
has twice more components than necessary and may not be used after
fixing the gauge. 
The 8-component supercharge must be constructed so that it would
reproduce the expression (45) for the $\kappa$-symmetry generator
in the space-time superalgebra.
The p-form terms of (45) should then appear as superpartners of
the p-form charges in the superalgebra.
Given these considerations,
the expression for the gauge-fixed supercharge should be written as: 
\eqn\grav{\eqalign{{Q^i_{SUSY-g.f.}}=
\oint{{dz}\over{2i\pi}}
\Gamma^{\bar{m}}_{{\bar\mu}{\bar\nu}}\Lambda^i_{\bar\mu}{\theta_{\bar\nu}}
\partial{X_{\bar{m}}}=\cr=
\oint{{dz}\over{2i\pi}}{\lbrace}
\Gamma_2{\partial}{\varphi^i}+{\Gamma_1}{\lbrack}\partial{\varphi^i}
(\Gamma^{{{\bar{m}}_1}...{{\bar{m}}_3}}\psi_{{\bar{m}}_1}...
\psi_{{\bar{m}}_3}+({\Gamma^{\bar{m}}}\psi_{\bar{m}})(\Gamma^{\bar{m}}
\partial{X_{\bar{m}}}))\cr
+\partial{X^i}(\Gamma^{{{\bar{m}}_1}...{{\bar{m}}_4}}\psi_{{\bar{m}}_1}...
\psi_{{\bar{m}}_4}+\Gamma^{{{\bar{m}}_1}{{\bar{m}}_2}}\psi_{{\bar{m}}_1}
\psi_{{\bar{m}}_2}(\Gamma^{\bar{m}}\partial{X_{\bar{m}}}))
{\rbrack}\cr
+e^{2\phi}{\lbrack}\partial{\varphi^i}(\Gamma^{{{\bar{m}}_1}...
{{\bar{m}}_8}}\psi_{{\bar{m}}_1}...\psi_{{\bar{m}}_8}+
\Gamma^{{{\bar{m}}_1}...{{\bar{m}}_6}}
\psi_{{\bar{m}}_1}...\psi_{{\bar{m}}_6}
(\Gamma^{\bar{m}}\partial{X_{\bar{m}}}))+\cr
+{\partial}X^i(\Gamma^{{{\bar{m}}_1}...{{\bar{m}}_7}}
\psi_{{\bar{m}}_1}...\psi_{{\bar{m}}_7}(\Gamma^{\bar{m}}\partial
{X_{\bar{m}}})+\Gamma^{{{\bar{m}}_1}...{{\bar{m}}_9}}
\psi_{{\bar{m}}_1}...\psi_{{\bar{m}}_9}){\rbrack}{\rbrace}}}
This expression for the space-time supersymmetry generator corresponds
to the spinor representation ${\bf{8_s}}$ of SO(8).
Beside that, we will need the expression for another gauge-fixed
supercharge in the space-time, corresponding to another
spinor representation ${\bf{8_c}}$ of SO(8):
\eqn\grav{\eqalign{{{\tilde{Q}}^j_{SUSY-g.f.}}=\oint{{dz}\over{2i\pi}}
\lbrace{\Gamma_2}{\varphi^j}+{\Gamma_1}{e^\phi}{\lbrack}{\varphi^j}
{\lbrack}{\Gamma^{{{\bar{m}}_1}
...{{\bar{m}}_3}}}\psi_{{\bar{m}}_1}...\psi_{{\bar{m}}_3}+
({\Gamma^{\bar{m}}}\psi_{\bar{m}})({\Gamma^{\bar{n}}}\partial{X_{\bar{n}}})
{\rbrack}+\cr+{\partial}{X^j}\Gamma^{{{\bar{m}}_1}{{\bar{m}}_2}}
\psi_{{\bar{m}}_1}\psi_{{\bar{m}}_2}{\rbrack}+\cr+
e^{2\phi}{\lbrack}{\varphi^j}{\lbrack}\Gamma^{{{\bar{m}}_1}...
{{\bar{m}}_8}}\psi_{{\bar{m}}_1}...\psi_{{\bar{m}}_8}+
\Gamma^{{{\bar{m}}_1}...{{\bar{m}}_6}}\psi_{{\bar{m}}_1}...
\psi_{{\bar{m}}_6}(\Gamma^{\bar{m}}\partial{X_{\bar{m}}}){\rbrack}
+\cr+\partial{X^j}{\lbrack}\Gamma^{{{\bar{m}}_1}...
{{\bar{m}}_7}}{\psi_{{\bar{m}}_1}}...\psi_{{\bar{m}}_7}+
\Gamma^{{{\bar{m}}_1}...{{\bar{m}}_5}}\psi_{{\bar{m}}_1}...
\psi_{{\bar{m}}_5}(\Gamma^{\bar{m}}\partial{X_{\bar{m}}}){\rbrack}
{\rbrace}}}
The appropriate anticommutator in the gauge-fixed  space-time
superalgebra will then be given by:
\eqn\lowen{{\lbrace}Q^i_{\bf{8s}},{\tilde{Q}}^j_{\bf{8c}}{\rbrace}=
\gamma^k_{ij}({\Gamma_4}\oint{{dz}\over{2i\pi}}(p^k+{\sum_{l=1}^9}
Z_{{{\bar{m}}_1}...{{\bar{m}}_l}}^k{\Gamma^{{{\bar{m}}_1}...
{{\bar{m}}_l}}}))}
where $p^k=:{\Gamma_1}e^{-\phi}{\varphi^k}:\sim{\partial}{X^k},k=1,...8$,
up to terms not contributing to correlation functions,
$\gamma^k$ are 8-dimensional gamma-matrices
and the p-form charges are given by:
\eqn\grav{\eqalign{
Z^k_{{\bar{m}}_1}={\Gamma_3}\oint{{dz}\over{2i\pi}}e^\phi{p^k}
{\psi_{{\bar{m}}_1}}(\Gamma^{{\bar{m}}}\partial{X_{\bar{m}}})\cr
Z^k_{{{\bar{m}}_1}{{\bar{m}}_2}}={\Gamma_3}\oint{{dz}\over2i\pi}
{e^\phi}{\varphi^k}\psi_{{\bar{m}}_1}\psi_{{\bar{m}}_2}
(\Gamma^{\bar{m}}\partial{X_{\bar{m}}})\cr
Z^k_{{{\bar{m}}_1}...{{\bar{m}}_3}}={\Gamma_3}\oint{{dz}\over{2i\pi}}
e^\phi{p^k}\psi_{{\bar{m}}_1}...\psi_{{\bar{m}}_3}\cr
Z^k_{{{\bar{m}}_1}...{{\bar{m}}_4}}={\Gamma_3}\oint{{dz}\over{2i\pi}}
e^\phi{\varphi^k}\psi_{{\bar{m}}_1}...\psi_{{\bar{m}}_4}\cr
Z^k_{{{\bar{m}}_1}...{{\bar{m}}_5}}={\Gamma_2}\oint{{dz}\over{2i\pi}}
e^{2\phi}{\varphi^k}{\partial}{X^{\bar{m}}}\partial{X_{\bar{m}}}
{\psi_{{\bar{m}}_1}}...{\psi_{{\bar{m}}_5}}\cr
Z^k_{{{\bar{m}}_1}...{{\bar{m}}_6}}={\Gamma_2}\oint{{dz}\over{2i\pi}}
e^{2\phi}{p^k}(\Gamma^{\bar{m}}\partial{X_{\bar{m}}})\psi_{{\bar{m}}_1}...
\psi_{{\bar{m}}_6}\cr
Z^k_{{{\bar{m}}_1}...{{\bar{m}}_7}}={\Gamma_2}\oint{{dz}\over{2i\pi}}
e^{2\phi}{\varphi^k}({\Gamma^{\bar{m}}}\partial{X_{\bar{m}}})
\psi_{{\bar{m}}_1}...\psi_{{\bar{m}}_7}\cr
Z^k_{{{\bar{m}}_1}...{{\bar{m}}_8}}={\Gamma_2}\oint{{dz}\over{2i\pi}}
e^{2\phi}{p^k}\psi_{{\bar{m}}_1}...\psi_{{\bar{m}}_8}\cr
Z^k_{{{\bar{m}}_1}...{{\bar{m}}_9}}={\Gamma_2}\oint{{dz}\over{2i\pi}}
e^{2\phi}{\varphi^k}{\psi_{{\bar{m}}_1}}...\psi_{{\bar{m}}_9}}}
Using these expressions for the p-form charges, 
taking (36),(45) into account and
deducing the $const$ of (43) to be equal to 1,
one may show that the
the $\kappa$-symmetry generator and, accordingly, the 
$\kappa$-invariance condition (25),(35) may be written in the form:
\eqn\lowen{{\oint}G^i_\kappa={\oint}\partial{\varphi^i}+\gamma^i_{kl}
{\lbrace}Q^k_{\bf{8s}},{\oint}p^l+\sum_{j=1}^{9}Z^l_{{{\bar{m}}_1}...
{{\bar{m}}_j}}\Gamma^{{{\bar{m}}_1}...{{\bar{m}}_j}}{\rbrace}=0}
 Therefore ,we see that the necessary condition for the $\kappa$-invariance,
the cancellation of the p-form terms, is related to the Bogomol'ny-type 
condition,applied to the anticommutator (51) of the
gauge-fixed space-time superalgebra,i.e. the vanishing of the r.h.s.
of the formula (51) which physically means the 
``charge = momentum'' condition, typical for the BPS saturation
criteria.
 This develops the arguments of the (47),applied to the situation
prior to the  gauge fixing and is in accordance with the above conjecture 
about the existing connection between the BPS saturation and the $\kappa$-
symmetry.We see that if the  p-brane solutions
of the D=10 supergravity (the low-energy limit of the GS 
superstring theory ) which are related to the p-form charges in the
superalgebra,
satisfy the BPS saturation condition,
one may hope that the $\kappa$-symmetry,formulated classically,
may also exist non-perturbatively.
 Given the fact that the $\kappa$-symmetry is crucial for the formulation
of space-time supersymmetric superstring and super p-brane theories,
its formulation on the non-perturbative level
would be important to the  understanding 
of the space-time supersymmetry in these theories,when the non-perturbative
effects are taken into account.In particular,one needs to develop
more precise and quantitive arguments that relate the p-form charges
in  superalgebras to p-branes.At present, the  question
of how this relation works, needs the further insight.

\centerline{\bf 7.Conclusion}
 
 In our approach the p-form charges and, in particular,the
RR charges of p-brane solutions to the low-energy effective theory,
appear as a result of the O.P.E.  between spin operators
for matter fields, contained in the NSR  formulations of both space-time
fermions $\theta_{\bar\mu}$ and the twistor fields 
$\Lambda_{\bar\mu}^i$.
In the previous paper ~\refs{\me} we have argued that,
in order to take the RR sector into account in the $\sigma$-model
approach,one had to supplement the standard $\sigma$-model action with
the worldsheet terms 
\eqn\lowen{I_{RR}\sim\int{d^2}{\tau}{\bar\Phi}F{\Phi}}
where $\bar\Phi$ and $\Phi$ are (0,1) and (1,0) matter-ghost
spin operators respectively and F are p-form RR charges, contracted
with gamma-matrices (with p - even and odd for the type IIA and IIB
theories respectively).Note that the spin operators $\Phi$ and
$\bar\Phi$ are 16-component ones, i.e. they are constucted so that
the local fermionic gauge symmetry still needs to be fixed.
Upon fixing the $kappa$-symmetry,the 16-component spinors
$\bar\Phi^{\bar\mu}$ and $\Phi^{\bar\mu}$
should be replaced with some 8-component fields,
corresponding to the spinor representations ${\bf{8_s}}$
or ${\bf{8_c}}$ of SO(8).
One may attempt to relate the gauge-fixed version of the p-form worldsheet
RR terms (54) to the p-form charges of (52),in other
words (52) should somehow arise from (54) in the process of fixing the gauge,
though we do not yet have convincing arguments to show that
and hope to do it in the future papers.
One should pay the special attention to the p-brane solutions
of the D=10 supergravity, coming from the RR sector,particularly,
given the fact that the p-brane solutions arising from the NS-NS sector
are either the elementary string
(which may be identified with the fundamental string), or
 the geodesically complete 5-brane solution ~\refs{\townsend}; therefore
the deep understanding of the RR sector is of the biggest interest.
Finally, let us make one concluding remark 
regarding the twistor-like approach.
Generalizing this method to the super fivebrane theory would lead
one to consider the theory with the N=1 supersymmetry on the 
$(5+1)$-volume (which would replace the $\kappa$-symmetry.Given the
established duality between string and fivebrane theories,
it would be interesting to try to study the relation between
the $N=1$ superalgebra on the (5+1)-volume and the $N=4$
superalgebra on the worldsheet,which  appeared previously
in our analysis of the string theory,supersymmetric in the
space-time. 
In general,exploring all the questions related to this subject,
one inevitably faces the problem of the consistent quantization
of super p-brane theories,which still waits  for the resolution.
\centerline{\bf Acknowledgements}
The author would like to express his sincere gratitude
to the Landau Institute for Theoretical Physics for the hospitality.

\listrefs

\end